\documentclass[runningheads]{llncs}

\usepackage{times}
\usepackage{enumerate}
\usepackage{xspace}
\usepackage{microtype}
\usepackage{proof}
\usepackage{amsfonts,bussproofs}

\newcommand{\egc}[0]{e.g.,\ }
\newcommand{\iec}[0]{i.e.,\ }

\newcommand{\commadots}[0]{,\ldots ,}

\def\n{\noindent}

\newcommand{\IMPL}[0]{\,\supset\,}

\newcommand{\dseq}[0]{\;\Rightarrow\;}
\renewcommand{\dseq}[0]{\Rightarrow}

\def\dt{T=\langle W,D\rangle}

\newcommand{\dtarg}[2]{\mbox{$\langle#1,#2\rangle$}}

\newcommand{\nop}[1]{}

\newcommand{\luk}{$\mathbf{\mbox{\textbf{\L}}_3}$\xspace}
\renewcommand{\luk}{${\mbox{\textbf{\L}}_\mathbf{3}}$\xspace}
\newcommand{\dlluk}{$\mathbf{\mbox{\textbf{D\L}}_3}$\xspace}

\renewcommand{\dlluk}{$\mathsf{DL}_{\mathsf{\mbox{\scriptsize \L}}_3}$}
\renewcommand{\dlluk}{$\mathbf{DL}_\mathbf{3}$\xspace}

\newcommand{\True}{\mathbf{t}}
\newcommand{\False}{\mathbf{f}}
\newcommand{\Indet}{\mathbf{u}}

\newcommand{\cert}[0]{\mathrm{L}\xspace} %
\newcommand{\poss}[0]{\mathrm{M}\xspace} %

\newcommand{\Thluk}[1]{\mathrm{Th}_{\,{\mbox{\scriptsize\textbf \L}}_\mathbf{3}}(#1)}

\newcommand{\val}{\ensuremath{\mathsf{v}}\xspace}

\newcommand{\rmid}{\ensuremath{\parallel}\space}
\renewcommand{\rmid}{\ensuremath{\nmid}\space}

\usepackage[T1]{fontenc}    
\usepackage[utf8]{inputenc} 
\usepackage{amsmath}
\usepackage{amssymb}
\usepackage{mathtools}
\usepackage{hyperref}
\begin{document}

\title{Sequent-Type Proof Systems for\\
Three-Valued Default Logic\thanks{I would like to thank my supervisor, Hans Tompits, for his indispensable help and guidance for all the work related to my master's thesis. Furthermore, I would like to thank the anonymous referees for valuable comments which helped to improve this extended abstract as well as for interesting suggestions for future work.
The partial support by the European Master's Program in Computational Logic (EMCL) is greatly acknowledged.}
}

\author{Sopo Pkhakadze}

\titlerunning{Sequent-Type Proof Systems for Three-Valued Default Logic}
\institute{Institute of Logic and Computation,\\
Knowledge-Based Systems Group E192-03,\\
Technische Universität Wien,\\
Favoritenstraße 9-11, 1040 Vienna, Austria \\
ORCID ID: 
{0000-0003-2247-8147}\\
\email{pkhakadze@kr.tuwien.ac.at}}

\maketitle 

\begin{abstract}
Sequent-type proof systems constitute an important and widely-used class of calculi well-suited for analysing proof search.
In my master's thesis, I introduce sequent-type calculi for a variant of default logic employing {\L}ukasiewicz's three-valued logic as the underlying base logic. This version of default logic has been introduced by Radzikowska addressing some representational shortcomings of standard default logic.
More specifically, the calculi discussed in my thesis axiomatise brave and skeptical reasoning for this version of default logic, respectively following the sequent method first introduced in the context of nonmonotonic reasoning by Bonatti and Olivetti, which employ a \emph{complementary calculus} for axiomatising invalid formulas, taking care of expressing the consistency condition of defaults.
\end{abstract}

\section{Background}

Nonmonotonic reasoning is a well-established area in knowledge representation and reasoning dealing with formalisations of \emph{rational arguments} whose characteristic feature is that their conclusions may have to be retracted in the light of new, more specific information.
Thus, the inference mechanism underlying rational arguments is \emph{nonmonotonic} in the sense that an increased set of premisses does not necessarily entail an increased set of conclusions.
This is in contradistinction to \emph{valid arguments} whose underlying inference process is \emph{monotonic}.
Many different nonmonotonic formalisms have been introduced in the literature, most prominent among them are default logic~\cite{reiter}, autoepistemic logic~\cite{Moore83}, circumscription~\cite{McCarthy:1980}, and logic programming under the answer-set semantics~\cite{Gelfond88,Gelfond91
}. 

In my thesis, I deal with a variant of default logic, viz.\ 
\emph{three-valued default logic}, \dlluk, introduced by Radzikowska~\cite{radzikowska96}, where 
{L}ukasiewicz's three-valued logic \luk~\cite{ukasiewicz19301} is used as underlying logical apparatus.
In particular, three-valued default logic allows for a more fine-grained distinction between formulas obtained by applying defaults and formulas which are known for certain, in order to avoid counterintuitive conclusions by successive applications of defaults.
Three-valued default logic is one of a variety of versions of default logic
addressing various shortcomings of the original proposal.
Among these different approaches are, \egc \emph{justified default logic}~\cite{lukasz88}, \emph{disjunctive default logic}~\cite{gel91a}, \emph{constrained default logic}~\cite{schaub92,delgrande-etal94}, \emph{rational default logic}~\cite{mikitiukT93}, and \emph{general default logic}~\cite{zhou-etal09} (an overview about different versions of default logic is given by Antoniou and Wang~\cite{antoniou-wang07}).

Similar as in standard default logic~\cite{reiter}, a \emph{default theory} in \dlluk is a pair $T=\dtarg{W}{D}$, where $W$ is a set of formulas in {\L}ukasiewicz's three-valued logic \luk and $D$ is a set of defaults which are rules of the form

\medskip
\centerline{
  \infer[,]
    {C}
    {A: B_{1} \commadots B_{n}}
    }
  
\n
where $A,B_{1},\ldots, B_{n},C$ are formulas in \luk.
The intuitive meaning of such a default is: 
\begin{quote}
if $A$ is believed, and $B_1\commadots B_n$ and $\cert C$ are consistent with what is believed (\iec, none of $\neg B_1\commadots \neg B_n, \neg\cert C$ are derivable), then $\poss C$ is asserted.  
\end{quote}
Here, 
$\cert$ and $\poss$ are operators which, according to {\L}ukasiewicz~\cite{ukasiewicz19301}, where first formalised in 1921 by Tarski by defining $\cert A := \neg(A\IMPL\neg A)$
and $\poss A := (\neg A\IMPL A)$.
Intuitively, $\cert A$ expresses that $A$ is \emph{certain}, whilst $\poss A$ means that $A$ is \emph{possible}.
With these operators, one distinguishes between \emph{certain knowledge} and \emph{defeasible conclusions}.
Note that \dlluk differs from the original version of default logic not only 
by using \luk instead of classical logic as the underlying logical apparatus 
but also by the modification of the consistency condition for applying defaults, having the additional condition that $\neg\cert C$ must also not be derivable.

\emph{Extensions}, representing a possible totality of logical consequences on the basis of a default theory, are defined by means of a fixed-point condition, similar as in standard default logic, but taking the modified interpretation of defaults and the underlying logic \luk into account.
Formally, an extension of a default theory $\dt$ in \dlluk is defined thus:
For a set $S$ of formulas, let $\Gamma_T (S)$ be the smallest set $K$ of formulas obeying
  the following conditions:
\begin{enumerate}[(i)]
\item
$K=\Thluk{K}$, where $\Thluk{K}$ is the \emph{deductive closure} of $K$ in \luk, \iec the set of all formulas derivable from $K$ in \luk;
\item
$W\subseteq K$;
\item
if $(A: B_{1},\ldots, B_{n}/ C)\in D$, $A\in K$, $\neg
B_{1}\not\in S,\ldots,\neg B_{n}\not\in 
S$, and $\neg\cert C\notin S$, then $\poss C\in K$.
\end{enumerate}
Then, $E$ is an extension of $T$ iff $\Gamma_T(E)=E$.


\section{Central Research Questions and Results}
In my thesis, I deal with the question of developing a proof theory for the three-valued default logic \dlluk based on the method of sequent-style calculi.
In general, 
sequent-type proof systems, first introduced in the 1930s by Gerhard Gentzen~\cite{Gentzen35} for classical and intuitionistic logic, are among the basic calculi used in automated deduction for analysing proof search.
Specifically, the aim of my thesis is to have systems axiomatising \emph{brave} and \emph{skeptical reasoning} for \dlluk.
Recall that a formula $A$ is a brave consequence of a default theory $T$ iff $A$ is contained in some extension of $T$, and $A$ is a skeptical consequence of $T$ iff it is contained in all extensions of $T$.

Although Radzikowska~\cite{radzikowska96} gave a resolution-based characterisation of brave reasoning for closed normal default theories, generalising the method for standard default logic as proposed by Reiter~\cite{reiter}, strictly speaking, this cannot count as a proper proof system because an external (meta-theoretical) consistency check has to be performed.
Rather, my thesis follows the approach first introduced by 
Bonatti~\cite{bonatti96} who developed a sequent calculus for propositional default logic, likewise
formalising brave reasoning as in Reiter's own proposal, but for general default theories
and not just normal ones.
Later, Bonatti and Olivetti~\cite{bonatti02} introduced also a calculus for skeptical reasoning and a variant calculus for brave reasoning.

In my thesis, the calculi of Bonatti~\cite{bonatti96} and Bonatti and Olivetti~\cite{bonatti02} are generalised to the three-valued default logic \dlluk of Radzikowska and soundness and completeness will be proven.
The elements of the brave reasoning calculus following the version of Bonatti~\cite{bonatti96} are sequents of the form 
$$\Gamma; \Delta \dseq \Sigma; \Theta,$$ where $\Gamma$, $\Sigma$, and $\Theta$ are finite sets of propositional formulas and $\Delta$ is a finite set of propositional defaults.
Such a sequent is true iff
  there is an extension $E$ of the default theory \dtarg{\Gamma}{\Delta} such
  that $\Sigma\subseteq E$ and $\Theta\cap E=\emptyset$.
Analogously, elements of the skeptical calculus are sequents of the form 
$$\Sigma; \Gamma; \Delta \dseq \Theta,$$ where $\Sigma$ is a set of formulas referred to as \emph{provability constraints}, $\Gamma$ and $\Theta$ are finite sets of propositional formulas, and $\Delta$ is a finite set of propositional defaults.
Such a sequent is true iff
all extensions of the default theory \dtarg{\Gamma}{\Delta} which satisfy the constraints in $\Sigma$, contain at least one element of $\Theta$.

A distinguishing feature of the calculi of Bonatti~\cite{bonatti96} and Bonatti and Olivetti~\cite{bonatti02} is the usage of a \emph{complementary calculus} for axiomatising invalid formulas, \iec of non-theorems, taking care of formalising the consistency condition of defaults, which makes these calculi arguably particularly elegant and suitable for proof-complexity elaborations as, \egc recently undertaken by Beyersdorff et al.~\cite{BeyersdorffMTV12}.
In a complementary calculus, the inference rules formalise the propagation
of refutability instead of validity and thus establish invalidity by deduction, \iec in a purely syntactic manner.
Complementary calculi are also referred to as \emph{refutation calculi} or \emph{rejection calculi} and the first axiomatic treatment of rejection was done by {\L}ukasiewicz in his formalisation of Aristotle's syllogistic~\cite{lukas39}.
Subsequently, rejection calculi for many different logics have been introduced, like for intuitionistic logic~\cite{skura89,dut89,skura99}, modal logics~\cite{goranko94,skura13}, many-valued logics~\cite{Tompits11,bogo14}, and description logics~\cite{berger-tompits13},
as well as a comprehensive theory of rejected propositions has been developed~\cite{slupecki59,wybr69,bryll69,bryll71,bryll72} (a detailed description of the history of axiomatic rejection is given, \egc in the excellent survey paper by Urszula Wybraniec-Skardowska~\cite{ursi05}).

Similar to Bonatti and Olivetti's approach, our calculi for \dlluk consist of three parts,~viz.\ 
\begin{enumerate}[(i)]
\item a sequent calculus for {\L}ukasiewiz's three-valued logic \luk, 
\item a complementary \emph{anti-sequent} calculus for \luk, and 
\item specific default inference rules.
\end{enumerate}
For many-valued logics, different kinds of sequent-style systems exist in the literature, like systems~\cite{beziau99,Avron02} based on (two-sided) sequents in the style of Gentzen~\cite{Gentzen35} employing additional non-standard rules, or using \emph{hypersequents}~\cite{avron91a}, which are tuples of Gentzen-style sequents.
In our sequent and anti-sequent calculi for {\L}ukasiewicz's three-valued logic \luk, we adopt the approach of Rousseau~\cite{Rousseau},
because it represents a natural generalisation of the classical two-sided sequent formulation of Gentzen to the many-valued case.
In a three-valued setting, a sequent in the sense of Rousseau is a triple of the form $$\Gamma_1\mid\Gamma_2\mid\Gamma_3,$$ where each $\Gamma_i$ ($i\in\{1,2,3\}$) is a finite set of formulas of \luk and each component of such a sequent intuitively corresponds to one of the three truth values in \luk, viz.\ $\Gamma_1$ corresponds to the truth value $\False$ (``false''), $\Gamma_2$ corresponds to  
$\Indet$ (``undetermined''), and $\Gamma_3$ corresponds to $\True$ (``true'').
More specifically, 
$\Gamma_1\mid\Gamma_2\mid\Gamma_3$ is true under a three-valued interpretation if, for at least one $i\in\{1,2,3\}$,  $\Gamma_i$ contains some formula $A$ having truth value $\val_i$, where $\val_1=\False$,   
$\val_2=\Indet$, and $\val_3=\True$, otherwise $\Gamma_1\mid\Gamma_2\mid\Gamma_3$ is false under the given interpretation.
Correspondingly, an anti-sequent is a triple of the form
$$\Gamma_1\rmid\Gamma_2\rmid\Gamma_3,$$ where each $\Gamma_i$ is as before, with the semantic meaning that $\Gamma_1\rmid\Gamma_2\rmid\Gamma_3$ is refutable iff $\Gamma_1\mid\Gamma_2\mid\Gamma_3$ is false under some three-valued interpretation.
Note that a three-valued interpretation is a mapping which assigns to each atomic formula of \luk one of the three truth values $\False$,   
$\Indet$, or $\True$, and this assignment of truth values is extended to arbitrary formulas by the respective truth conditions for the different logical connectives of \luk (for details, cf., \egc the well-known textbook by Malinowski~\cite{malinowski93} or the survey paper by the same author~\cite{malinowski07}).

The calculi which will be used for three-valued sequents and anti-sequents can be obtained from a systematic construction of calculi for many-valued logics as described by Zach~\cite{Zach93} and Bogojeski~\cite{bogo14}.

\section{Discussion}

The proof-theoretical approach we undertake is flexible and can be applied to formalise also other versions of default reasoning.
Indeed, justified default logic~\cite{lukasz88}, constrained default logic~\cite{schaub92,delgrande-etal94}, and rational default logic~\cite{mikitiukT93} have been axiomatised in the style of Bonatti~\cite{bonatti96} by Lupea~\cite{lupea08}, as well as intuitionistic default logic by Egly and Tompits~\cite{EglyT97}.

Related to the sequent approach discussed in my thesis are works employing tableau methods.
In particular, Niemelä~\cite{niemela-ecai96} introduces a tableau calculus for inference under circumscription.
Other tableau approaches, however, \emph{do not encode inference directly}, rather they characterise \emph{models} (resp., \emph{extensions}) associated with a particular nonmonotonic reasoning formalism~\cite{Amati:1996,PearceGV00,CabalarOPV07,GebserS13}.

In view of the close relation of standard default logic with answer-set programming, an interesting possible topic for future work would be to study a similar relation of \dlluk to a {\L}ukasiewicz-style three-valued semantics  of answer-set programs.
Note that the concept of \emph{Kleene answer-set programs}~\cite{DohertyS14,DohertyKS16,DohertyS16} have recently been defined, making use of the three-valued logic of Kleene~\cite{kleene38}.
Also, as \dlluk allows to distinguish between certain knowledge and default conclusions, one may envisage a mechanism to keep track on how many applications of defaults are required in order to derive a certain default conclusion, and thus being able to define some kind of degree of trust of a default conclusion, as done similarly in the approach of Rondogiannis and Troumpoukis~\cite{RondogiannisT13} for logic programs under the well-founded semantics.

\section{Publications}

The calculus for brave reasoning for \dlluk generalising Bonatti's approach~\cite{bonatti96} from my thesis have already been published in the proceedings of the 15th International Conference on Logic Programming and
Nonmonotonic Reasoning (LPNMR 2019), while the prospective results concerning the other calculi are planned to be submitted to a future conference.


\end{document}